\documentclass[aps,apl,twocolumn,floatfix,superscriptaddress]{revtex4}
\usepackage{graphicx}
\usepackage{amsmath}
\usepackage{hyperref}
\usepackage{bm}
\usepackage[dvips]{color}

\newcommand{\degr}{$^{\circ}$}
\newcommand{\etal} {\textit{et al.}}
\newcommand{\ie} {\textit{i.e.}}
\newcommand {\eg} {\textit{e.g.}}

\newcommand{\moss} {M\"ossbauer}
\newcommand{\fzd} {Institute of Ion Beam Physics and Materials Research, Forschungszentrum Dresden-Rossendorf, P.O. Box 510119, 01314 Dresden, Germany}

\newcommand{\chim} {$\chi_{min}$}

\begin{document}
\title{Crystallographically oriented Co and Ni nanocrystals inside ZnO formed by ion implantation and post-annealing}

\date{\today}

\author{Shengqiang~Zhou}
\email[Electronic address: ]{S.Zhou@fzd.de}
\author{K.~Potzger}
\author{J.~von Borany}
\author{R.~Gr\"{o}tzschel}
\author{W.~Skorupa}
\author{M.~Helm}
\author{J.~Fassbender}
\affiliation{\fzd}

\begin{abstract}

In the last decade, transition metal doped ZnO has been
intensively investigated as a route to room temperature diluted
magnetic semiconductors (DMS). However the origin for the reported
ferromagnetism in ZnO based DMS remains questionable. Possible
options are diluted magnetic semiconductors, spinodal
decomposition or secondary phases. In order to clarify this
question, we have performed a thorough characterization of the
structural and magnetic properties of Co and Ni implanted ZnO
single crystals. Our measurements reveal that Co or Ni
nanocrystals (NCs) are the major contribution of the measured
ferromagnetism. Already in the as-implanted samples, Co or Ni NCs
have formed, and they exhibit superparamagnetic properties. The Co
or Ni NCs are crystallographically oriented with respect to the
ZnO matrix. Their magnetic properties, \eg~the anisotropy and the
superparamagnetic blocking temperature can be tuned by annealing.
We discuss the magnetic anisotropy of Ni NCs embedded in ZnO
concerning the strain anisotropy.

\end{abstract}
\maketitle

\section{Introduction}\label{introduction}

Recently diluted magnetic semiconductors (DMS) are under intensive
investigation due to their potential applications in spintronics.
In DMS materials, transition or rare-earth metal ions are
substituted onto cation sites and are coupled with free carriers
to yield ferromagnetism via indirect interaction. Mn-doped InAs
\cite{munekataInMnAs} and GaAs \cite{ohnoGaMnAs} are the best
studied DMS materials. Conventional III-V semiconductors are
widely used for high-speed electronic and optoelectronic devices.
The discovery of hole-mediated ferromagnetism in (Ga,Mn)As opened
the way to integrate spin-based phenomena into mainstream
microelectronics and optoelectronics as well as taking advantage
of already established fabrication processes. Electrically
controlled spintronic devices based on GaMnAs and InMnAs have been
successfully designed and tested at low temperatures, \eg, a
spin-LED using GaMnAs as a spin injector \cite{ohnoSPINLED}.
However the highest Curie temperature (T$_C$) reported in
(Ga,Mn)As is $\sim$170 K \cite{edmonds05}, which is far below room
temperature and limits its regular application.

In 2000, Dietl \etal~\cite{dietl00} used a mean field theory to
estimate the ordering temperature T$_C$ of ferromagnetic
semiconductors, and they predict that room-temperature
ferromagnetism can be created by substituting Mn ions in p-type
wide-band gap semiconductors such as GaN and ZnO. Sato
\etal~calculated the properties of n-type ZnO doped with 3$d$ TM
ions (V, Cr, Mn, Fe, Co, and Ni) \cite{sato_ZnO}. The
ferromagnetic state, with a T$_C$ of around 2000 K, is predicted
to be favourable for V, Cr, Fe, Co, and Ni in ZnO while Mn-doped
ZnO is predicted to be antiferromagnetic. These predictions
largely boosted intensive experimental activities on transition
metal doped GaN and ZnO. A large number of research groups have
reported the experimental observation of ferromagnetism in TM
(from Sc to Ni) doped ZnO
\cite{angadi06,heo04,hong05,hongv,ip03,jung02,pol04,tuan04,venk04}
fabricated by various methods including ion implantation. For a
comprehensive review, see Ref. \cite{pearton03m,
PeartonPMS,PeartonIEEE}. However in these reports the magnetic
properties using the same dopant vary considerably.
\textit{E.g.}~the saturation moment and Curie temperature for Mn
doped ZnO ranges from 0.075$\mu_B$/Mn, 400 K \cite{hong05} to
0.17$\mu_B$/Mn, 30-45 K \cite{jung02}, respectively. In contrast
to these publications, other groups reported the observation of
antiferromagnetism \cite{boul05,yin06,sati:137204}, spin-glass
behavior \cite{fukumura01,jin01}, and paramagnetism
\cite{yin06,rao05,zhang06,zhou07V} in TM-doped ZnO. Recently it
was also found that nanoscale precipitates can contribute to the
ferromagnetic properties substantially
\cite{norton03,pot06fe,zhou06,zhou07JPD,zhou07si,talut06,jamet06,liu:092507}.

Publications claiming the intrinsic ferromagnetism in TM doped ZnO
are often based solely on magnetization measurements using high
sensitivity superconducting quantum interference device
magnetometery (SQUID) and structure characterizations using
lab-equipped x-ray diffraction. The latter has been demonstrated
to be not sensitive enough to detect nanoscale precipitates
\cite{pot06fe}. Nowadays the research community puts more effort
to judge if TM ions in semiconductors are homogenously
distributed, and if the ferromagnetism is intrinsic
\cite{dietlmat,kuroda07}. Anomalous Hall effect (AHE), that
verifies charge carrier participation in the magnetic order, has
been suggested to judge about carrier-mediated ferromagnetism
arising from DMS. However a recent study presents the
co-occurrence of superparamagnetism and AHE in Co doped TiO$_2$
films \cite{shinde04}. Another criterium is the magnetic
anisotropy \cite{venk04,sati:017203,leejw07} as a signature of the
intrinsic ferromagnetism. The controversy in the magnetic
properties of ZnO-based DMS, as stated above, might partially be
due to the insufficient characterization of the samples
\cite{fukumura04,liu05,seschadri05}. Particularly, a careful
correlation between structure and magnetism should be established
by sophisticated methods. Synchrotron radiation based x-ray
diffraction (SR-XRD) is a powerful tool to detect small
precipitates, \eg~metallic TM nanocrystals in ZnO \cite{pot06fe}.
On the other hand, element selective measurements of the magnetic
properties, \eg~XMCD \cite{kim03}, and \moss~spectroscopy
\cite{pot06fe,talut06}, address the origin of the measured
magnetism directly.

The present work is devoted to a comprehensive investigation of Co
and Ni doped ZnO. By SR-XRD and SQUID magnetometry, we correlate
the structural and magnetic properties. We attempt to answer the
following questions concerning the presence of ferromagnetic
precipitates.

(1) How do the crystalline precipitates orient with respect
    to the host matrix?

(2) Which techniques are suitable to detect these
    precipitates?

(3) Are these precipitates the major contribution of the
    measured ferromagnetism, or is there another source which
    contributes?

(5) Can these nanoscale precipitates exhibit magnetic
properties,\eg~magnetic anisotropy, concerning the criteria above
mentioned?

(6) How do these precipitates behave upon thermal annealing?

The paper is organized as follows. In section \ref{experiment} all
the experimental methods employed will be described. Then the
results will be presented in two sections. In section
\ref{as_implanted} we will focus on the as-implanted samples, and
discuss the orientation, the superparamagnetism, the magnetic
anisotropy of Co and Ni NCs. In section \ref{post_annealing} we
will describe the structure and magnetism evolution due to high
vacuum annealing. Finally in section \ref{discussion} we discuss
the origin of the magnetic-anisotropy for the oriented Co, Ni NCs
system, and the possible formation of Co/CoO and Ni/NiO core/shell
structures upon annealing at 923 K. This paper is concluded in
section \ref{conclusion}.

\section{Experiments}\label{experiment}
Commercial ZnO single crystals grown by the hydrothermal method
were implanted with Co or Ni ions at 623 K with a fluence ranging
from $0.8\times10^{16}$ cm$^{-2}$ up to $8\times10^{16}$
cm$^{-2}$. The implantation energy was 180 keV, which resulted in
a projected range of $R_P=89\pm29$ nm, and a maximum atomic
concentration from $\sim$1\% to $\sim$10\% (TRIM code
\cite{trim}). Thermal annealing was performed in a high vacuum
($<$10$^{-6}$ mbar) furnace from 823 K to 1073 K.

The lattice damage induced by implantation was evaluated by
Rutherford backscattering/channeling spectrometry (RBS/C). RBS/C
spectra were collected with a collimated 1.7 MeV He$^+$ beam at a
backscattering angle of 170\degr. The sample was mounted on a
three-axis goniometer with a precision of 0.01\degr. During
channeling measurement, the sample was aligned to make the
ZnO$<$0001$>$ axis parallel to the impinging He$^+$ beam. \chim,
the channeling minimum yield in RBS/C, is the ratio of the
backscattering yield at channeling condition to that for a random
beam incidence \cite{chuwk}. Therefore, \chim~labels the degree of
lattice disorder upon implantation, and an amorphous sample shows
a \chim~of 100\%, while a perfect single crystal corresponds to a
\chim~of 1-2\%.

Magnetic properties were measured with a superconducting quantum
interference device (SQUID, Quantum Design MPMS) magnetometery.
The samples were measured with the field along both the in- and
out-of-plane direction. We studied both the temperature dependence
of the magnetization at a constant field and the field dependence
at a constant temperature. By SQUID, virgin ZnO is found to be
purely diamagnetic with a susceptibility of -2.65$\times$10$^{-7}$
emu/Oe$\cdot$g. This background has been subtracted from the
magnetic data. The temperature dependent magnetization measurement
has been carried out in the following way. The sample was cooled
in zero field from above room temperature to 5 K. Then a 50 Oe
field was applied, and the zero field cooled magnetization curve
(ZFC curve) was measured with increasing temperature from 5 to 350
K, after which the field cooled magnetization curve (FC curve) was
measured in the same field from 350 to 5 K with decreasing
temperature.

Structural analysis was performed both by synchrotron radiation
x-ray diffraction (SR-XRD) and conventional XRD. SR-XRD was
performed at the Rossendorf beamline (BM20) at the ESRF with an
x-ray wavelength of 0.154 nm. Conventional XRD was performed with
a Siemens D5005 equipped with a Cu-target source. In XRD
measurement, we use 2$\theta$-$\theta$ scans to identify
crystalline precipitates, and pole figures (azimuthal $\phi$-scan)
for determining their crystallographical orientation. As a
standard approach, for an XRD $\phi$-scan one first tilts the
sample by the angle of $\chi$ from the sample surface ($\chi$ is
the angle between the diffraction plane of interest and the sample
surface), and fixes the Bragg angle. Subsequently the spectrum is
recorded during azimuthal rotation with respect to the sample
normal. The pole figure is constructed by a series of $\phi$-scans
at different $\chi$.

\section{As-implanted samples}\label{as_implanted}

\subsection{ZnO lattice damage upon implantation}

\begin{figure} \center
\includegraphics[scale=0.80]{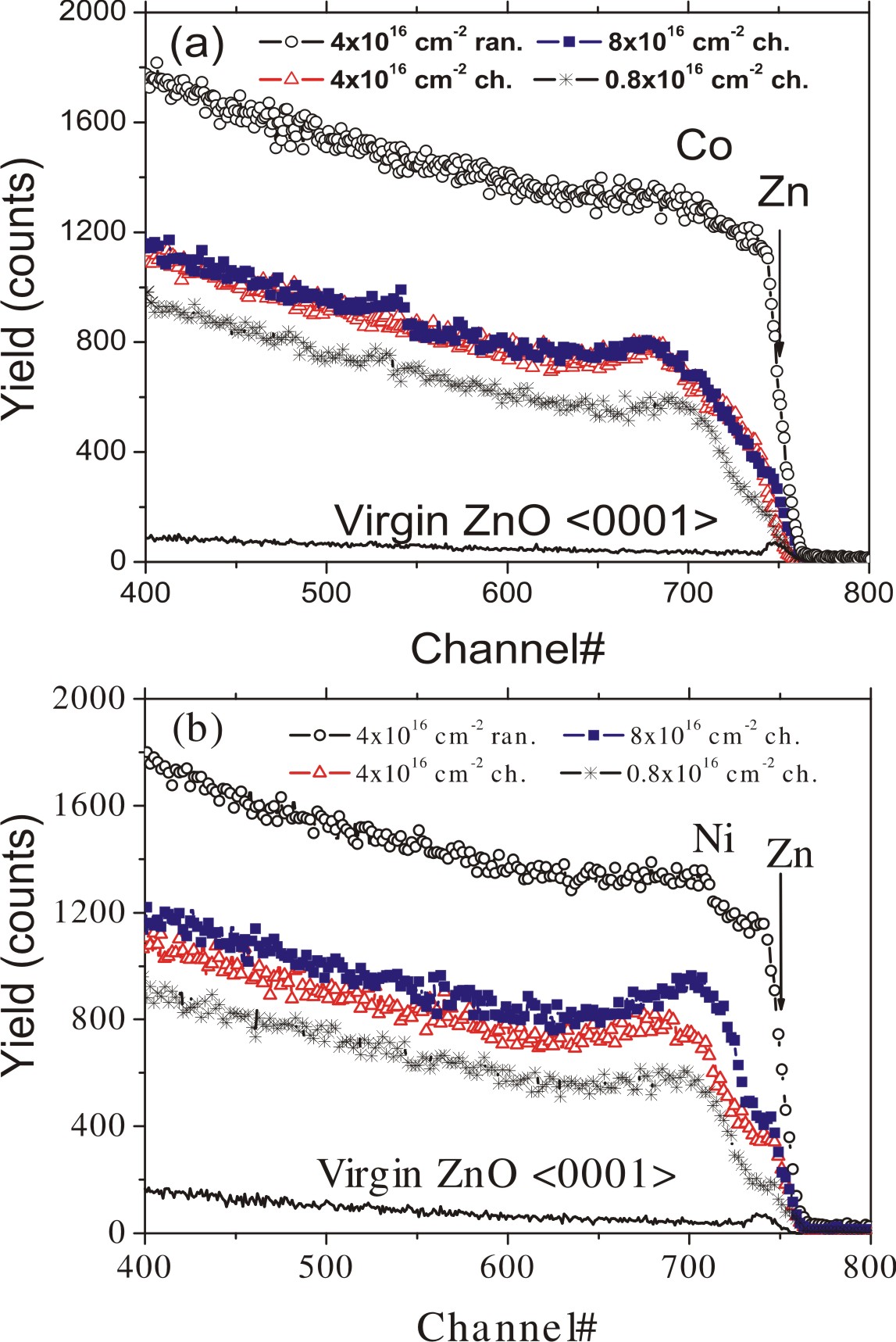}
\caption{RBS random (ran.) and channeling (ch.) spectra, (a) Co
implanted ZnO, and (b) Ni implanted ZnO (The fluence for Co and Ni
ions is indicated). The yield of channeling spectra is
progressively decreased with increasing
fluence.}\label{fig:RBS_CoNiZnO_asimp}
\end{figure}

RBS/C is used to check the lattice damage after implantation.
Figure \ref{fig:RBS_CoNiZnO_asimp} shows RBS/C spectra for
different fluences. The arrow labelled Zn indicates the energy for
backscattering from surface Zn atoms. The implanted Co or Ni ions
cannot be detected for the very low fluence (0.8$\times$10$^{16}$
cm$^{-2}$, not shown). However, they are more pronounced as a hump
in the random spectrum for a larger fluence of 4$\times$10$^{16}$
cm$^{-2}$ and 8$\times$10$^{16}$ cm$^{-2}$ (not shown). The humps
in the channeling spectra mainly originate from the lattice
disordering due to implantation. As expected, \chim~increases with
increasing fluence (see Table \ref{tab:CoNiZnO_asimp}). Note that
the highest Co fluence induced comparable lattice damage with the
middle fluence, and less than the damage created by the same
fluence of Ni. The reason could be a dopant specific
self-annealing process \cite{kucheyev03}. RBS/C measurements also
reveal that the ZnO is a hard material with respect to
irradiation. The host material still partly remains in a
crystalline state after irradiation with Co and Ni ions up to a
fluence of 8$\times$10$^{16}$ cm$^{-2}$ (\chim~of 59\% and 69\%,
respectively).

\subsection{Crystalographically oriented Co and Ni NCs} \label{section:XRD_asimp}
\begin{figure} \center
\includegraphics[scale=0.8]{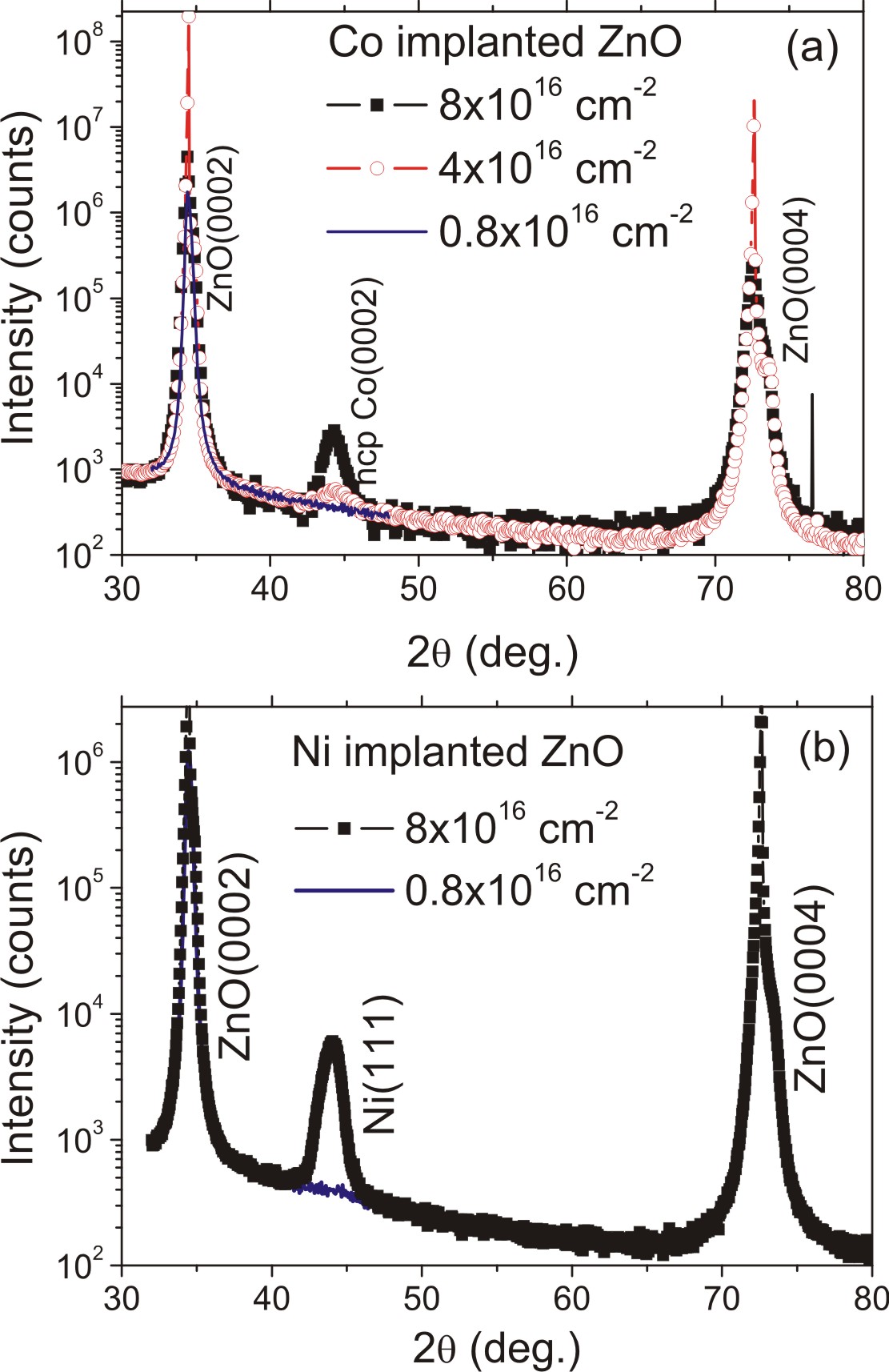}
\caption{SR-XRD 2$\theta$-$\theta$ scan revealing the existence of
Co or Ni precipitates in (a) Co implanted ZnO, and (b) Ni
implanted ZnO.
}\label{fig:XRD_CoNiZnO_asimp}
\end{figure}

SR-XRD is used to identify the precipitates in ZnO after Co or Ni
implantation. Figure \ref{fig:XRD_CoNiZnO_asimp}~shows the XRD
$2\theta$-$\theta$ scans for all samples implanted with different
fluences. At a low fluence (0.8$\times$10$^{16}$ cm$^{-2}$), no
crystalline Co or Ni NCs could be detected. At a fluence of
4$\times$10$^{16}$ cm$^{-2}$ the hcp-Co(0002) (or fcc-Co(111)) and
Ni(111) peak appear, respectively, and grow with increasing
fluence. The full width at half maximum (FWHM) of the Co or Ni
peak decreased with fluence, indicating the growth of the average
diameter of these NCs (table \ref{tab:CoNiZnO_asimp}). The
crystallite size is calculated using the Scherrer formula
\cite{scherrer},

\begin{equation}\label{scherrer}
    d=0.9\lambda/(\beta\cdot\cos\theta),
\end{equation} where $\lambda$ is the wavelength of the x-ray, $\theta$ the bragg angle, and $\beta$ the FWHM of
2$\theta$ in radians.

Note that there is only one peak for Co or Ni detectable, which
indicates that the crystallites of Co or Ni are highly oriented
with respect to the host matrix. The surface orientation is
hcp-Co(0001)(or fcc-Co(111))$\parallel$ZnO(0001) and
fcc-Ni(111)$\parallel$ZnO(0001), respectively.

\begin{figure} \center
\includegraphics[scale=0.56]{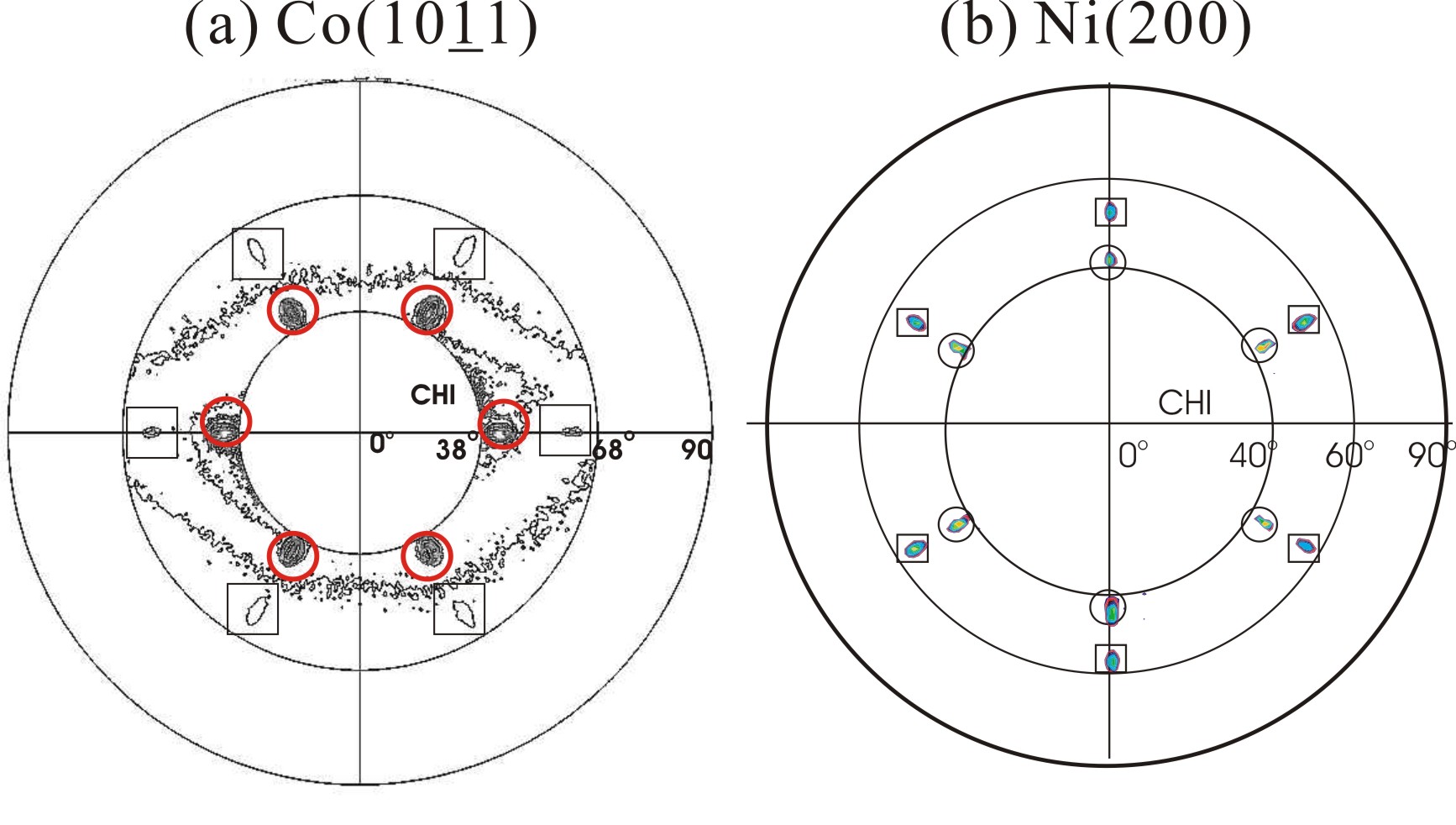}
\caption{XRD Pole figure revealing the crystallographical
orientation relationship between Co/Ni NCs and ZnO matrix, (a)
Co(10$\underline{1}$1) (in square) together with the tail of
ZnO(10$\underline{1}$2) (in circle); The data points out of
squares and circles are due to the background; (b) Ni(200) (in
square) together with the tail of ZnO(10$\underline{1}$2) (in
circle).}\label{fig:CoNiZnO_polefigure}
\end{figure}

The Co hcp structure only differs in the stacking from the fcc
one. Since the Bragg angles ($\theta$) for hcp-Co(0002)
($\theta$=22.38\degr) and fcc-Co(111) ($\theta$=22.12\degr) are
rather close to each other, it is difficult to assign these peaks
in Figure \ref{fig:XRD_CoNiZnO_asimp}(a) to hcp-Co or fcc-Co. A
$\phi$-scan or pole figure on one of the diffraction planes not
parallel with the sample surface (i.e. tilted by an angle $\chi$
from sample surface) helps to identify hcp or fcc-Co NCs and also
reveals the crystallographical orientation relationship. By this
approach, we find that only hcp-Co is present in the as-implanted
samples. Figure \ref{fig:CoNiZnO_polefigure} (a) and (b) show the
pole figure of hcp-Co(10$\underline{1}$1) and Ni(200),
respectively. The radial coordinate is the angle ($\chi$) by which
the surface is tilted out of the diffraction plane. The azimuthal
coordinate ($\phi$) is the angle of rotation about the surface
normal. The pole figure shows poles of hcp-Co(10$\underline{1}$1)
at $\chi$$\sim$61.9\degr, and Ni(200) at $\chi$$\sim$54.8\degr,
respectively. Both exhibit a sixfold symmetry. Since
ZnO(10$\underline{1}$2) and hcp-Co(10$\underline{1}$1) have
similar Bragg angle, the poles of ZnO(10$\underline{1}$2) also
show up at $\chi$$\sim$42.8\degr~with much more intensities. The
results are consistent with the theoretical Co(10$\underline{1}$1)
pole figure viewed along [0001], and Ni(200) pole figure viewed
along [111] direction, respectively. Therefore, we can conclude
that these Co and Ni NCs are crystallographically oriented with
respect to the ZnO matrix. The in-plane orientation relationship
is hcp-Co[10$\underline{1}$0]$\parallel$ZnO[10$\underline{1}$0],
and Ni[112]$\parallel$ZnO[10$\underline{1}$0], respectively. Due
to the hexagonal structure of Co and sixfold symmetry of Ni viewed
along [111] direction, it is not difficult to understand their
crystallographical orientation onto hexagonal-ZnO.

At this point, we have to remind the reader that in our former
work on Ni implanted ZnO, the XRD measurement was performed on a
conventional lab-equipped diffractometry (CXRD) \cite{zhou06}.
CXRD reveals similar results as SR-XRD (from the Ni fluence of
4$\times$10$^{16}$ cm$^{-2}$, Ni NCs start to form). However, CXRD
fails to detect Fe NCs in ZnO \cite{pot06fe}, where Fe NCs are not
well oriented like the case of Ni in ZnO.  The peak intensity in
XRD is proportional to the diffraction volume, \ie~the number of
diffraction planes which are parallel to each other. For a
2$\theta$-$\theta$ scan, the crystallographic orientation results
in much more diffraction volume than the random orientation.
Therefore, the high ordered orientation of NCs make them easier to
be detected. As expected, Co NCs are also detectable in the CXRD
measurement (not shown).

\begin{table*}
\caption{\label{tab:CoNiZnO_asimp}Structural and magnetic
properties for Co and Ni-implanted ZnO with different fluence.
Metallic Co/Ni fraction corresponds to the percentage of metallic
Co/Ni compared with all implanted Co/Ni.}
\begin{ruledtabular}
\begin{tabular}{cccccccc}
  Fluence  &  \chim~ & Crystallite size & Saturation magnetization\footnotemark[1] & Metallic & Coercivity\footnotemark[1] & T$_{B}$ & Crystallite size\footnotemark[2]\\
  (cm$^{-2}$) & (RBS/C) & (nm) (XRD) & ($\mu_B$/Co or /Ni) & fraction & (Oe) & (K) & (nm) \\
  \hline
  Co: 0.8$\times$10$^{16}$ & 44\% & - & - & - & - & - & \\
  Co: 4$\times$10$^{16}$ & 54\% & 5 & 0.29 (5 K) & 17\% & 1400 (5 K) & 45 & 4.3 \\
  Co: 8$\times$10$^{16}$ & 57\% & 8 & 0.44  (5 K) & 26\% & 1400 (5 K) & 300 & 8.1 \\
  \\
  Ni: 0.8$\times$10$^{16}$ & 45\% & - & 0.05 (10 K) & 8\% & 10 (10 K) & $\leq$ 5 & $<$ 9.4 \\
  Ni: 4$\times$10$^{16}$ & 57\% & 6 & 0.16 (10 K) & 27\% & 30 (10 K) & 16 & 14 \\
  Ni: 8$\times$10$^{16}$ & 69\% & 8 & 0.22 (10 K) & 37\% & 120 (10 K) & 70 & 23 \\
\end{tabular}
\end{ruledtabular}
\footnotetext[1]{Refer to the easy axis at a saturation field of
10000 Oe for Co-ZnO, and 1500 Oe for Ni-ZnO,
respectively.}\footnotetext[2]{Calculation from the average
blocking temperature by Eq. \ref{blocking_squid}.}
\end{table*}

\subsection{Magnetic properties of Co and Ni NCs}

The magnetic properties of Co and Ni implanted ZnO were measured
by SQUID magnetometery with the field parallel and perpendicular
to the sample surface. Co implanted samples exhibit a hard axis
parallel to the sample surface, while the easy axis is
perpendicular to the surface. For Ni-implanted samples, the
anisotropy directions are vice versa. In this section, we
investigate the superparamagnetism of the implanted ZnO samples
and their magnetic anisotropy.

\subsubsection{Superparamagnetic Co and Ni NCs}

\begin{figure} \center
\includegraphics[scale=0.80]{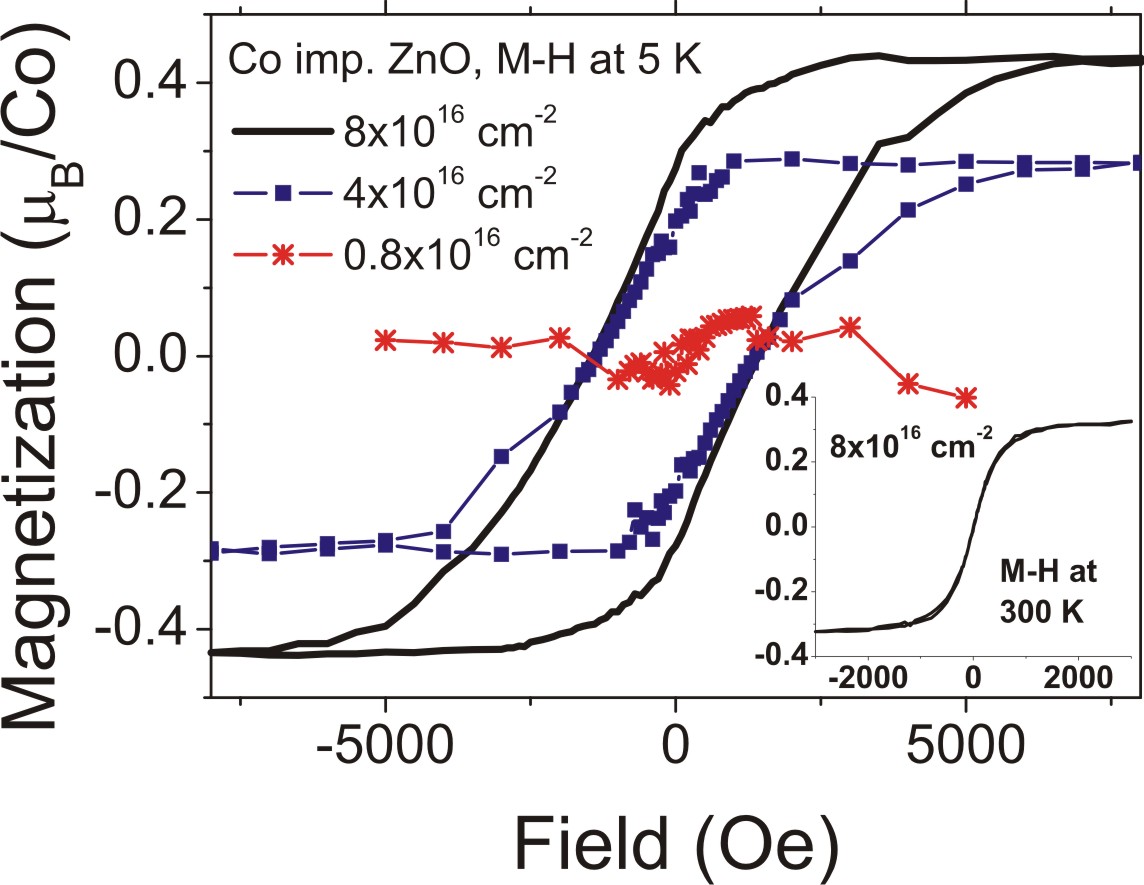}
\caption{Hysteresis loops measured at 5 K for Co implanted ZnO
with different fluences. Inset: Hysteresis loop measured at 300 K
for the sample with the highest
fluence.}\label{fig:CoZnO_MH_asimp}
\end{figure}

From the XRD results, we know that Co and Ni NCs have been formed
in the as-implanted samples. For magnetic nanoparticles, the
formation of domain walls is energetically unfavorable. Below a
certain size (typically in the range of 15 to 30 nm depending on
the material), the particle stays in a single-domain
configuration. If the particle size is sufficiently small, above a
particular temperature (so-called blocking temperature, T$_B$)
thermal fluctuations dominate and no preferred magnetization
direction can be defined. Such a system of superparamagnetic
particles does not exhibit a hysteresis curve above T$_B$;
therefore the coercivity (H$_C$) and the remanence (M$_R$) are
both zero.

For a dc magnetization measurement in a small magnetic field by
SQUID, T$_{B}$ is given by
\begin{equation}\label{blocking_squid}
    T_{B,Squid}\approx\frac{K_{eff}V}{30k_B},
\end{equation} where $K_{eff}$ is the anisotropy energy density, $V$ the particle volume, $k_B$ the Boltzmann constant \cite{respaud}. For bulk crystals, $K_{eff}(V)$ is 5.0$\times$10$^5$ and 5.7$\times$10$^3$ Jm$^{-3}$ for Co and Ni, respectively, at room temperature.

Phenomenologically there are two characteristic features in the
temperature dependent magnetization of a nanoparticle system. One
is the irreversibility of the magnetization in a small applied
field (\eg~50 Oe) after zero field cooling and field cooling
(ZFC/FC) \cite{respaud}. The other is the drastic drop of
coercivity and remanence at a temperature close to or above T$_B$
\cite{shinde04}.

Figure \ref{fig:CoZnO_MH_asimp} shows the magnetization versus
field reversal (M-H) measured at 5 K with the field applied
perpendicular to the sample surface (along ZnO[0001]). A
hysteretic behavior is observed for the high-fluence implanted
samples. A saturation behavior is also observed at 300 K for the
sample with the highest fluence (Figure \ref{fig:CoZnO_MH_asimp}
inset). However, neither coercivity nor remanence can be observed
at 300 K. This is a strong indication for the superparamagnetism
of a magnetic nanoparticle system. Knowing the formation of hcp-Co
from XRD, it is reasonable to conclude that hcp-Co NCs are
responsible for the magnetic behavior. For bulk hcp-Co crystals,
the magnetic moment is 1.7 $\mu_{B}$/Co at 0 K. Assuming the same
value for Co NCs, around 17\% and 26\% of implanted Co ions are in
the metallic state for the fluence of 4$\times$ and
8$\times$10$^{16}$ cm$^{-2}$ respectively. Similar results are
observed for Ni implanted ZnO \cite{zhou06}.

Note that the hysteresis loop for the fluence of
4$\times$10$^{16}$ cm$^{-2}$ exhibits two reversal steps. There is
a very small kink when the field decreases from -2000 to -4000 Oe
(and increases from 2000 to 4000 Oe). This wasp-waist shape of the
loop is associated with magnetic phases with different
coercivities \cite{zhou07si,brem:224427}.

\begin{figure} \center
\includegraphics[scale=0.8]{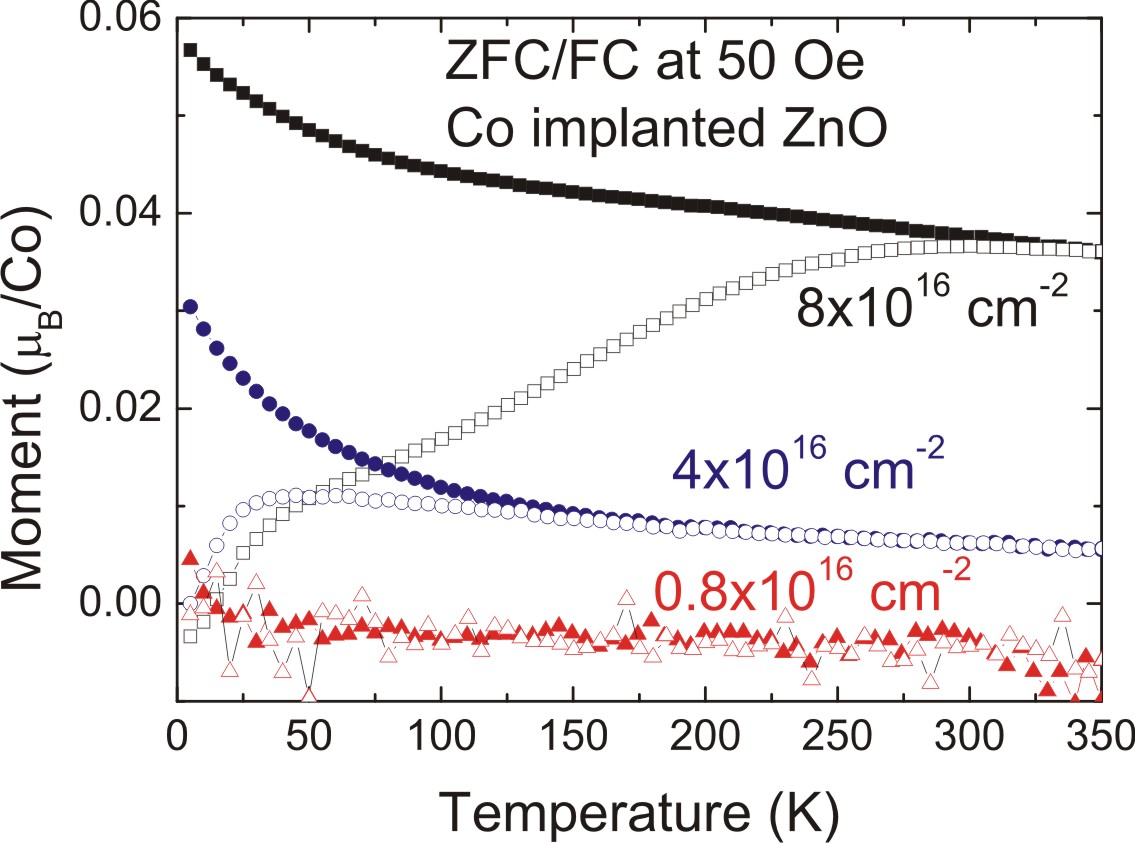}
\caption{Magnetization curves at 50 Oe after ZFC/FC for the Co
implanted ZnO. With increasing fluence, the Co NCs is growing in
size, resulting in a higher blocking
temperature.}\label{fig:CoZnO_ZFCFC_asimp}
\end{figure}

Temperature dependent magnetization with H$=50$ Oe was measured
after ZFC and FC to confirm the superparamagnetism (Figure
\ref{fig:CoZnO_ZFCFC_asimp}). For the magnetic samples, a distinct
difference in ZFC/FC curves is observed. ZFC curves show a gradual
increase (deblocking) at low temperature, and reach a broad peak,
while FC curves continue to increase with decreasing temperature.
The broad peaks in ZFC curves are due to the size distribution of
Co NCs. In this paper, the temperature at the maximum of the ZFC
curve is taken as the average blocking temperature (later referred
as T$_B$). The ZFC/FC curves are general characteristics of
magnetic nanoparticle systems with a broad size distribution
\cite{tsoi:014445}. T$_B$ increases with the fluence, \ie~the size
of nanoparticles. Table \ref{tab:CoNiZnO_asimp}~lists the average
size of Co NCs calculated by Eq. \ref{scherrer} (XRD) and Eq.
\ref{blocking_squid} (SQUID). They are in a good agreement
although there is also a size distribution in Co NCs \cite{farle}.
The ZFC/FC magnetization was also measured for Ni implanted
samples \cite{zhou06}. Comparing with Co, Ni has a much lower
anisotropy energy density. For similar sizes of Ni NCs, the
blocking temperature is therefore much lower than that of Co.
Table \ref{tab:CoNiZnO_asimp}~lists the average size of Ni NCs
calculated by Eq. \ref{scherrer} (XRD) and Eq.
\ref{blocking_squid} (SQUID). Although the trend is the same for
both calculations, the values from SQUID are much larger than that
from XRD data. One reason could be the anisotropy energy density
is underestimated by assuming the magneto-crystalline anisotropy
constant. Another reason is that T$_B$ is overestimated by taking
the temperature at the maximum of the ZFC curve. According to the
calculation by Farle \etal~\cite{farle}, and by Jacobsohn
\etal~\cite{jacobsohn:321}, the size dispersion for a given
average particle diameter broadens the peak of ZFC curve and
shifts it to higher temperatures.

\subsubsection{Magnetic anisotropy}

\begin{figure} \center
\includegraphics[scale=0.45]{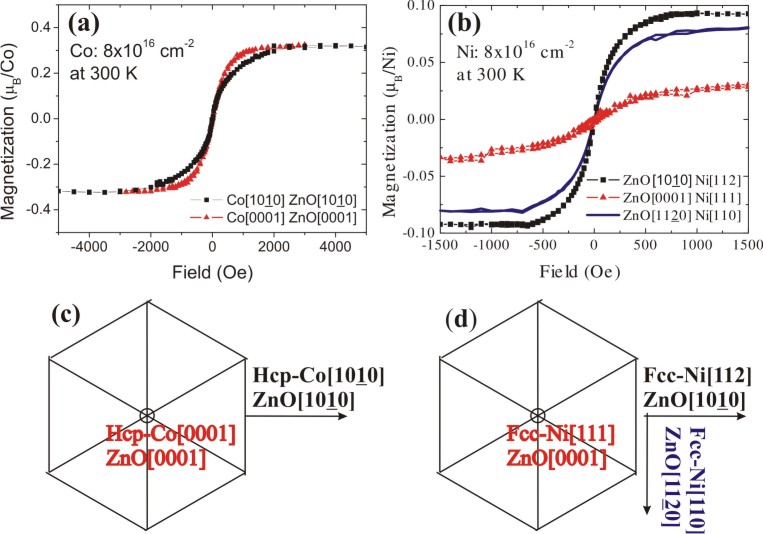}
\caption{Hysteresis loops measured with the field along ZnO[0001]
(out-of-plane) and [10$\underline{1}$0] (in-plane) for Co/Ni
implanted ZnO with the fluence of 8$\times$10$^{16}$ cm$^{-2}$
measured at 300 K, (a) Co implanted ZnO; and (b) Ni implanted ZnO.
(c) and (d) show the schematic geometry for magnetization
measurements.}\label{fig:CoNiZnO_aniso_asimp}
\end{figure}

M-H loops were also measured for selective samples with Co fluence
of 8$\times$10$^{16}$ cm$^{-2}$ with the field both perpendicular
and parallel to the sample surface. Figure
\ref{fig:CoNiZnO_aniso_asimp}(a) shows the comparison of the
magnetization along ZnO[10$\underline{1}$0] and [0001] at 300 K.
Figure \ref{fig:CoNiZnO_aniso_asimp}(c) shows the orientation
relationship between hcp-Co and ZnO, and the measurement geometry.
Obviously Co[0001] is the easy axis, the same as a bulk hcp-Co
crystal. The intersection of both curves gives an effective
anisotropy field of 3000 Oe. At 5 K, the magnetic anisotropy is
the same (not shown), and the coercivity of the easy axis is
around 1400 Oe. The ratio between remanence and saturation
magnetization is around 60\%. This rather low remanence (below
100\%) is due to the size distribution of nanomagnets. Very small
nanomagnets behave superparamagnetically even at low temperatures,
and only have field induced magnetization. This is a rather
universal feature for nanomagnets. For instance, epitaxial MnAs
nanoclusters in GaAs, the remanence is also below 100\% along the
easy axis \cite{ando:387}. Figure \ref{fig:CoNiZnO_aniso_asimp}(b)
shows the same measurement of Ni implanted ZnO, while (d) shows
the orientation relationship between fcc-Ni and ZnO, and the
measurement geometry. In contrast to bulk Ni where [111] is the
easy axis, the easy axis is Ni[112] and the hard axis is Ni[111].
Moreover, as shown in Figure \ref{fig:CoNiZnO_aniso_asimp}(b),
another in-plane direction Ni[110] is also an easy axis. Within
the applied field, the magnetization curve along hard axis does
not intersect with that along the easy axis. The effective
anisotropy field is much larger than 1500 Oe. That means that
there are other contributions to the anisotropy dominating over
the crystalline magnetic anisotropy. This will be discussed in
section \ref{discussion}.

In the work by Norton \etal~\cite{norton03}, epitaxial Co
nanocrystals have been observed in Co implanted ZnO single
crystals. The nanocrystal size is estimated to be $\sim$3.5 nm,
which is below the superparamagnetic limit at room temperature.
Therefore the ferromagnetism above 300 K is very possible due to
Co substitution onto the Zn site in the ZnO matrix. In our case,
the measured superparamagnetism is well explained by the presence
of Co and Ni nanocrystals. The formation of metallic nanocrystals
already in the as-implanted state is due to the elevated
implantation temperature (623 K), which facilities the
precipitation process. Implantation at low temperatures (\eg~253
K) prohibits precipitation, however results in non-magnetism in
our case, \eg~Fe implanted ZnO \cite{zhouFe}.

\section{The effect of Post-annealing}\label{post_annealing}

The post-annealing was performed in high vacuum with temperatures
ranging from 823 K to 1073 K for 15 min. The samples to be
annealed were selected according to the SQUID measurement
capability, namely the temperature range from 1.8 to 400 K. Due to
the higher anisotropy energy of Co, T$_B$ in ZFC curve of the
highest fluence is already 300 K in the as-implanted state (see
Figure \ref{fig:CoZnO_ZFCFC_asimp}). The annealing is expected to
increase the size of Co NCs, and consequently increase T$_B$ in
the ZFC curve, which will exceed the temperature range of the
SQUID magnetometery. Therefore, the sample implanted with the
middle Co fluence was chosen. For Ni the highest implantation
fluence was chosen.

\subsection{Lattice recovery}

As shown in Figure \ref{fig:RBS_CoNiZnO_asimp}, ion implantation
substantially induces lattice damage of ZnO crystals. Here we show
that this damage can be partially recovered by post-annealing.

\begin{figure} \center
\includegraphics[scale=0.80]{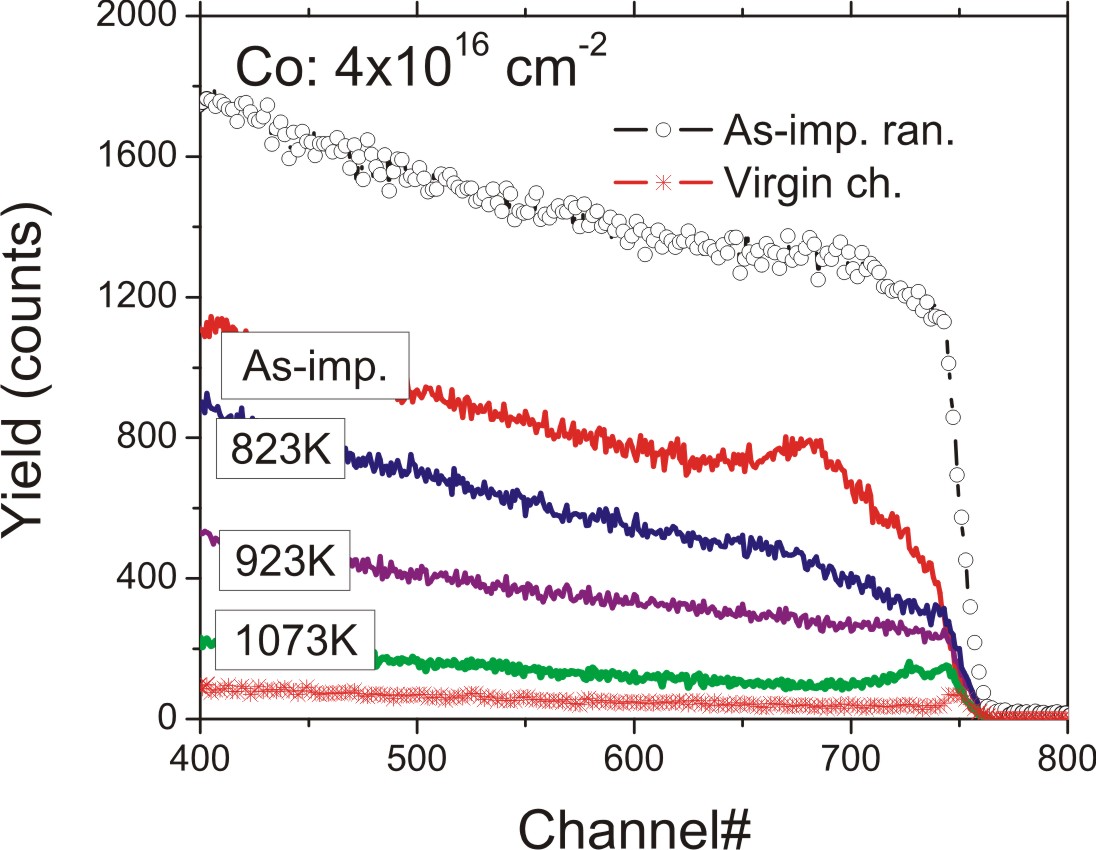}
\caption{RBS random (ran.) and channeling (ch.) spectra for Co
implanted ZnO with a fluence of 4$\times$10$^{16}$ cm$^{-2}$ after
thermal annealing at different temperatures. The lattice damage
induced by implantation is progressively reduced by increasing
annealing temperature.}\label{fig:RBS_CoZnO_ann}
\end{figure}

Figure \ref{fig:RBS_CoZnO_ann} shows the RBS/C spectra for the
sample implanted with a Co fluence of 4$\times$10$^{16}$
cm$^{-2}$. The annealing temperatures are given in the figure.
With increasing annealing temperature, the channeling spectra
indicate that the lattice disorder of ZnO progressively decreases.
After annealing to 1073 K, the channeling spectrum is almost
comparable with the virgin ZnO. Similar RBS/C results are observed
for Ni-implanted ZnO upon annealing (not shown).

\subsection{Evolution of structural properties}

\begin{figure} \center
\includegraphics[scale=0.80]{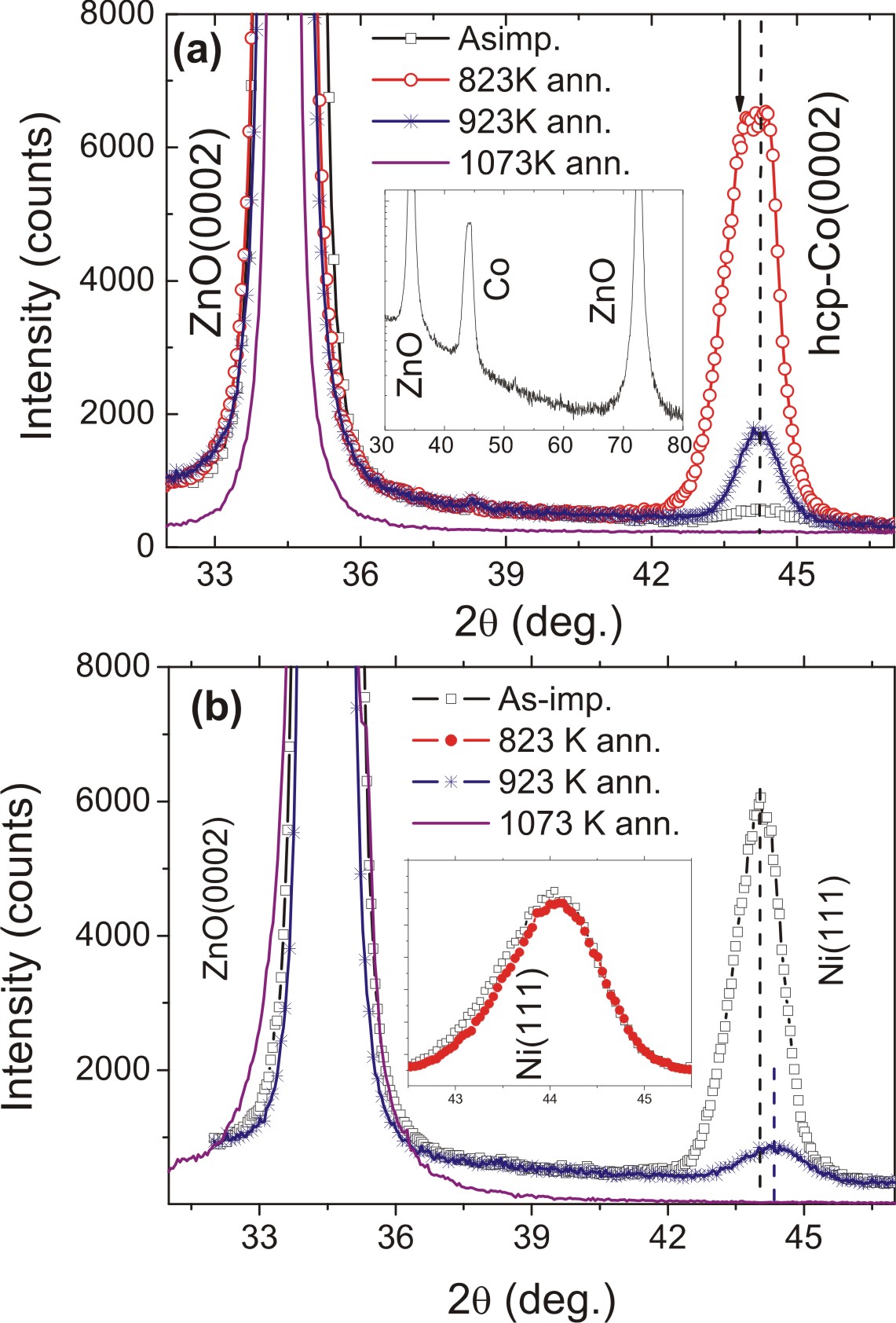}
\caption{XRD 2$\theta$-$\theta$ scans: (a) Co implanted ZnO
crystals with different annealing temperature. The wide range XRD
pattern for one of the samples (inset) reveals that no other
crystalline phase (e.g. CoO) could be detected. The arrow points
the peak shoulder coming from fcc-Co(111) diffraction in the
sample of 823 K ann. (b) Ni implanted ZnO crystals with different
annealing temperature. The inset shows a comparison between
as-implanted and 823 K annealed
samples.}\label{fig:XRD_CoNiZnO_ann}
\end{figure}

\begin{figure} \center
\includegraphics[scale=0.80]{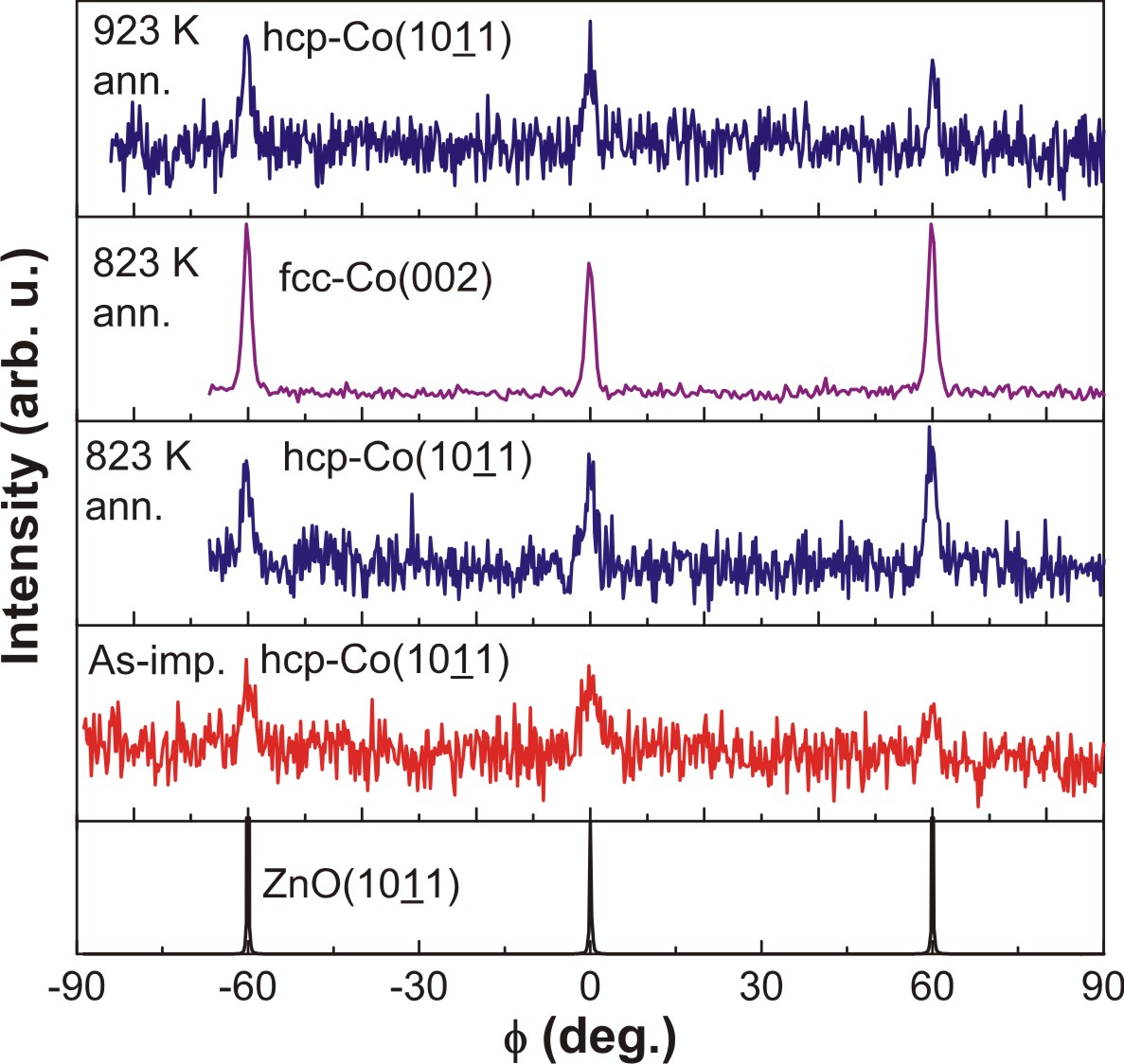}
\caption{XRD $\phi$-scans for hcp-Co(10$\underline{1}$1)
($\theta$=23.78\degr~and $\chi$$\sim$61.9\degr), fcc-Co(002)
($\theta$=25.76\degr~and $\chi$$\sim$54.8\degr) and
ZnO(10$\underline{1}$1) ($\theta$=18.13\degr~and
$\chi$$\sim$61.6\degr) reveal the in-plane orient relationship for
Co NCs respect to ZnO.}\label{fig:phi_CoZnO}
\end{figure}

\begin{figure} \center
\includegraphics[scale=0.9]{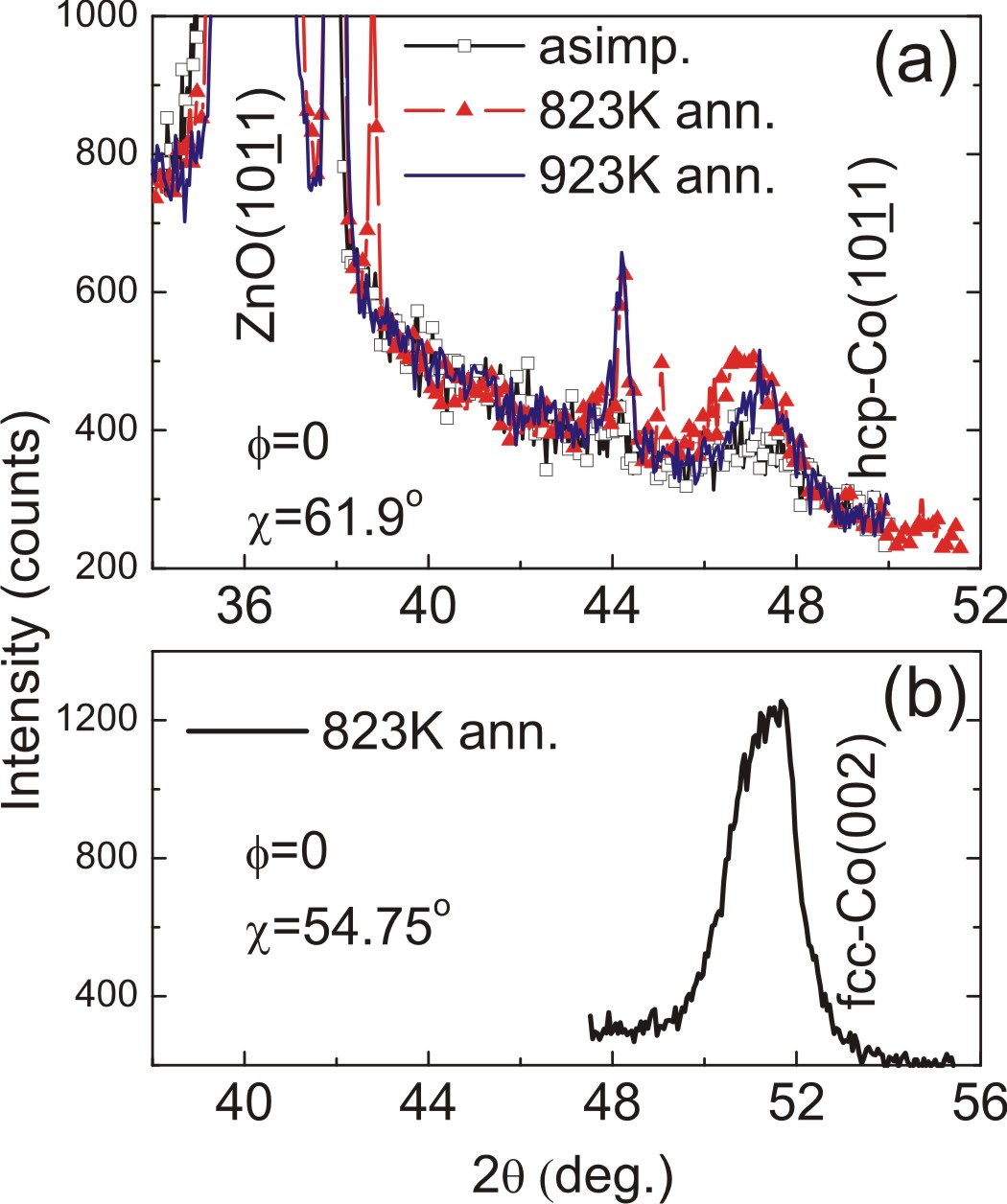}
\caption{(a) 2$\theta$-$\theta$ scans for ZnO(10$\underline{1}$1)
and hcp-Co(10$\underline{1}$1), those small sharp peaks are from
artificial noise in the measurements; (b) 2$\theta$-$\theta$ scan
for fcc-Co(002).}\label{fig:Co_skew}
\end{figure}

Figure \ref{fig:XRD_CoNiZnO_ann}(a) shows the development of Co
NCs upon thermal annealing. The peak area and crystallite size
calculated using the Scherrer formula \cite{scherrer} are compared
in table I. A broad scan (the inset of Figure
\ref{fig:XRD_CoNiZnO_ann}(a)) reveals only one peak from Co
besides the ZnO peaks. An XRD $\phi$-scan has been used to
distinguish between hcp- and fcc-Co (Figure \ref{fig:phi_CoZnO}).
We find only hcp-Co in the as-implanted sample and the sample
annealed at 923 K, while both fcc- and hcp-Co are present in the
sample annealed at 823 K. The broad peak in Figure
\ref{fig:XRD_CoNiZnO_ann}(a) (823 K ann.) is a superposition of
hcp-Co(0002) and fcc-Co(111). The crystallographical orientation
relationship between Co NCs and ZnO is
hcp-Co(0001)[10$\underline{1}$0]//ZnO(0001)[10$\underline{1}$0]//fcc-Co(111)[112].
With this orientation, the 2$\theta$-$\theta$ scans for
hcp-Co(10$\underline{1}$1) and fcc-Co(200) are expected with a
skew geometry at one of the azimuthal position (\eg~at $\phi$=0)
as shown in Figure \ref{fig:Co_skew}(a) and (b), respectively. In
skew geometry, the incident and the diffracted waves have the same
angles to the surface, while the sample is titled with respect to
its surface normal. By this configuration, a noncoplanar, its
surface normal does not lie in the plane defined by the incident
and the diffracted waves, can be measured \cite{kaganer:045423}.
Note that the peak area of Co in Figure \ref{fig:XRD_CoNiZnO_ann},
which is an approximate measure of the amount of Co NCs, increases
drastically after 823 K annealing, while decreases after 923 K
annealing. It is reasonable to attribute this change to the
formation and disappearance of fcc-Co. The fcc-Co is probably
oxidized to the amorphous CoO after 923 K annealing, while finally
the majority of Co NCs are oxidized to amorphous state after
annealing at 1073 K.

For Ni implanted ZnO, the structure evolution upon annealing is
similar as shown in Figure \ref{fig:XRD_CoNiZnO_ann}(b). The peak
area and the crystallite size is listed in table
\ref{tab:CoNiZnO_ann}. The mild temperature annealing (823 K) only
slightly increases the grain size of Ni. The annealing at 923 K
drastically decreases the peak area, while the grain size also
decreases. Note that there is a significant shift in the Ni(111)
peak (labelled by dash lines), which indicates the relexation of
lattice strain. After annealing at 1073 K, the majority of Ni NCs
could be oxidized to amorphous state.

\subsection{Evolution of magnetic properties}

\begin{table*}
\caption{\label{tab:CoNiZnO_ann}Structural and magnetic properties
for Co and Ni-implanted ZnO with different fluence. Metallic Co/Ni
fraction corresponds to the percentage of metallic Co/Ni compared
with all implanted Co/Ni.}
\begin{ruledtabular}
\begin{tabular}{cccccccc}
  Sample  &  Peak area & Crystallite size & Saturation magnetization\footnotemark[1] & Metallic & Coercivity\footnotemark[1] & T$_{B}$ & Crystallite size\footnotemark[2]\\
   & (XRD) & (nm) (XRD) & ($\mu$$_B$/Co or /Ni) & fraction & (Oe) & (K) & (nm) \\
  \hline
  Co: As-imp. & 380 & 5 & 0.29 (at 10000 Oe) & 17\% & 1400 & 45 & 4.3 \\
  Co: 823 K ann. & 8330 & - & 0.36 (at 2000 Oe)  & 21\% & 250 & 80 & 5.2 \\
  Co: 923 K ann. & 1400 & 10 & 0.32 (at 2000 Oe) & 19\% & 450 & 330 & 8.3 \\
  Co: 1073 K ann. & 0 & - & - & - & - & - & -\\
  \\
  Ni: As-imp. & 7312 & 8 & 0.22 (at 2000 Oe) & 37\% & 230 & 70 & $<$ 9.4 \\
  Ni: 823 K ann. & 6590 & 9 & 0.22 (at 2000 Oe) & 37\% & 200 & 80 & 14 \\
  Ni: 923 K ann. & 750 & 7 & 0.18 (at 2000 Oe) & 30\% & 220 & - & - \\
  Ni: 1073 K ann. & 0 & - & - & - & - & - & - \\
\end{tabular}
\end{ruledtabular}
\footnotetext[1]{Refer to the easy axis magnetization at 5 K at a
saturation field as indicated.}\footnotetext[2]{Calculation from
the average blocking temperature by Eq. \ref{blocking_squid}.}
\end{table*}

\begin{figure} \center
\includegraphics[scale=0.44]{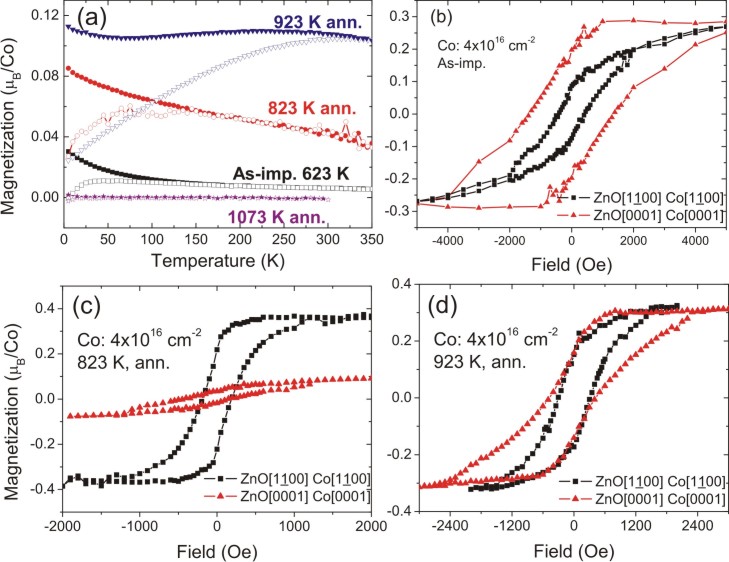}
\caption{(a) ZFC/FC magnetization curves at 50 Oe for the samples
after implantation and annealing at different temperatures. Solid
symbols are FC curves, while open symbols are ZFC curves; (b)-(d)
M-H curves measured at 5 K for all samples: along
ZnO[10$\underline{1}$0]$\parallel$hcp-Co[10$\underline{1}$0]$\parallel$fcc-Co[112]
(solid symbols) and
ZnO[0001]$\parallel$hcp-Co[0001]$\parallel$fcc-Co[111] (open
symbols). Implantation or annealing temperature is
shown.}\label{fig:SQUID_CoZnO_ann}
\end{figure}

The structural phase transformation of Co NCs results in different
magnetic properties as revealed by SQUID. Figure
\ref{fig:SQUID_CoZnO_ann}(a) shows the ZFC/FC magnetization curves
for all samples annealed at different temperatures. Obviously,
except the sample annealed at 1073 K, the ZFC curves show a
gradual increase (deblocking) at low temperature, and reach a
maximum at a temperature of T$_B$ (shown in table
\ref{tab:CoNiZnO_ann}), while FC curves continue to increase with
decreasing temperature. No significant magnetization response is
detected for the sample annealed at 1073 K. Note that T$_B$
increases drastically above 330 K after annealing at 923 K.
However Jacobsohn \etal~\cite{Jacobsohn} reported a much lower
T$_B$ (250 K) of hcp Co NCs with similar grain size ($\sim$10 nm).
The higher T$_B$ could be due to some other anisotropy energy,
which stablized the superparamagnetism at higher temperature. This
will be discussed in section \ref{discussion}.

\begin{figure} \center
\includegraphics[scale=0.46]{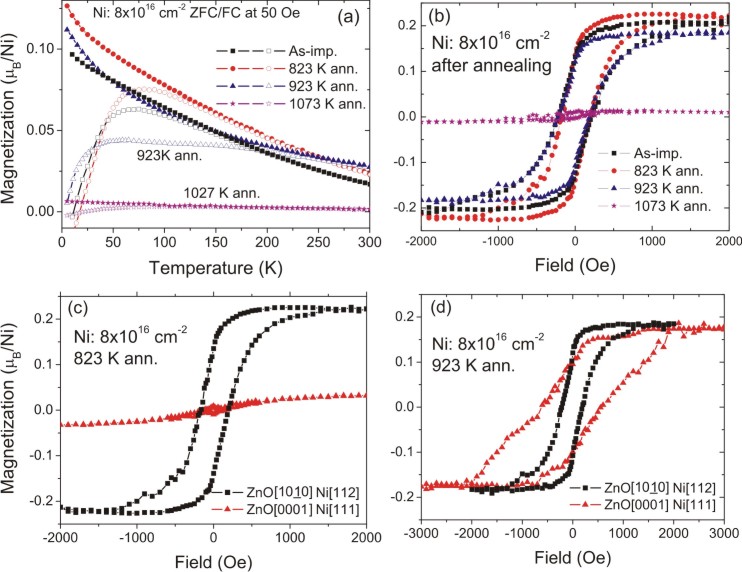}
\caption{(a) ZFC/FC magnetization curves at 50 Oe for the samples
after implantation and annealing at different temperatures. Solid
symbols are FC curves, while open symbols are ZFC curves; (b) M-H
curves measured at 5 K for all samples along
ZnO[10$\underline{1}$0]$\parallel$fcc-Ni[112]; (c) and (d) M-H
curves measured at 5 K for all samples along
ZnO[10$\underline{1}$0]$\parallel$fcc-Ni[112] (solid symbols) and
ZnO[0001]$\parallel$fcc-Co[111] (open symbols) after 823 K and 923
K annealing, respectively. Implantation or annealing temperature
is shown.}\label{fig:SQUID_NiZnO_ann}
\end{figure}

Hysteresis loops were measured for all samples in both parallel
(ZnO[10$\underline{1}$0]$\parallel$hcp-Co[10$\underline{1}$0]$\parallel$fcc-Co[112])
and perpendicular
(ZnO[0001]$\parallel$hcp-Co[0001]$\parallel$fcc-Co[111])
directions. Figures \ref{fig:SQUID_CoZnO_ann}(b)-(d) reveal that
the anisotropy and coercivity can be tuned by different annealing
procedures. The as-implanted sample only consists of hcp Co NCs,
which persist the bulk-like anisotropy along [0001] direction with
a high coercivity (Figure \ref{fig:SQUID_CoZnO_ann}(b)). The
sample annealed at 823 K mainly consists of fcc-Co with easy axis
is along the fcc-Co[112] direction (Figure
\ref{fig:SQUID_CoZnO_ann}(c)). After annealing at 923 K, the
sample shows an easy axis both along hcp-Co[10$\underline{1}$0]
and hcp-Co[0001], but a higher coercivity along the latter
direction. The magnetic properties of different samples are listed
in table \ref{tab:CoNiZnO_ann}.

Figure \ref{fig:SQUID_NiZnO_ann} show the magnetic properties of
Ni implanted ZnO upon annealing. As expected, the 823 K annealing
increases T$_B$ in the ZFC curve due to the increase of
crystallite size. However annealing at 923 K, the shape of ZFC
magnetization curve is deviated from others. There is no real
maximum, but a broad plateau. The formation of Ni/NiO core/shell
structure could introduce the exchange coupling, which contributes
another anisotropy energy. This will be discussed in the following
section. Note that after annealing at 1073 K the sample still
exhibits a non-neglectful response in ZFC/FC magnetization
measurement. There are could be a small amount of Ni nanocrystals
remaining, while they are beyond the detection limit in SR-XRD
measurement.

\section{Discussion}\label{discussion}

\subsection{Magnetic anisotropy of Co and Ni NCs}

As shown before, the magnetic anisotropy of Ni NCs embedded in ZnO
are drastically different from the bulk Ni. For bulk Ni, the
magnetocrystalline anisotropy constant of K$_1$ is
-5.7$\times$10$^3$ J m$^{-3}$ at 300 K. The [111] direction is the
easy axis. For a single magnetic NC, the magnetocrystalline,
shape, and magnetoelastic anisotropy have to be considered.

\begin{figure} \center
\includegraphics[scale=0.60]{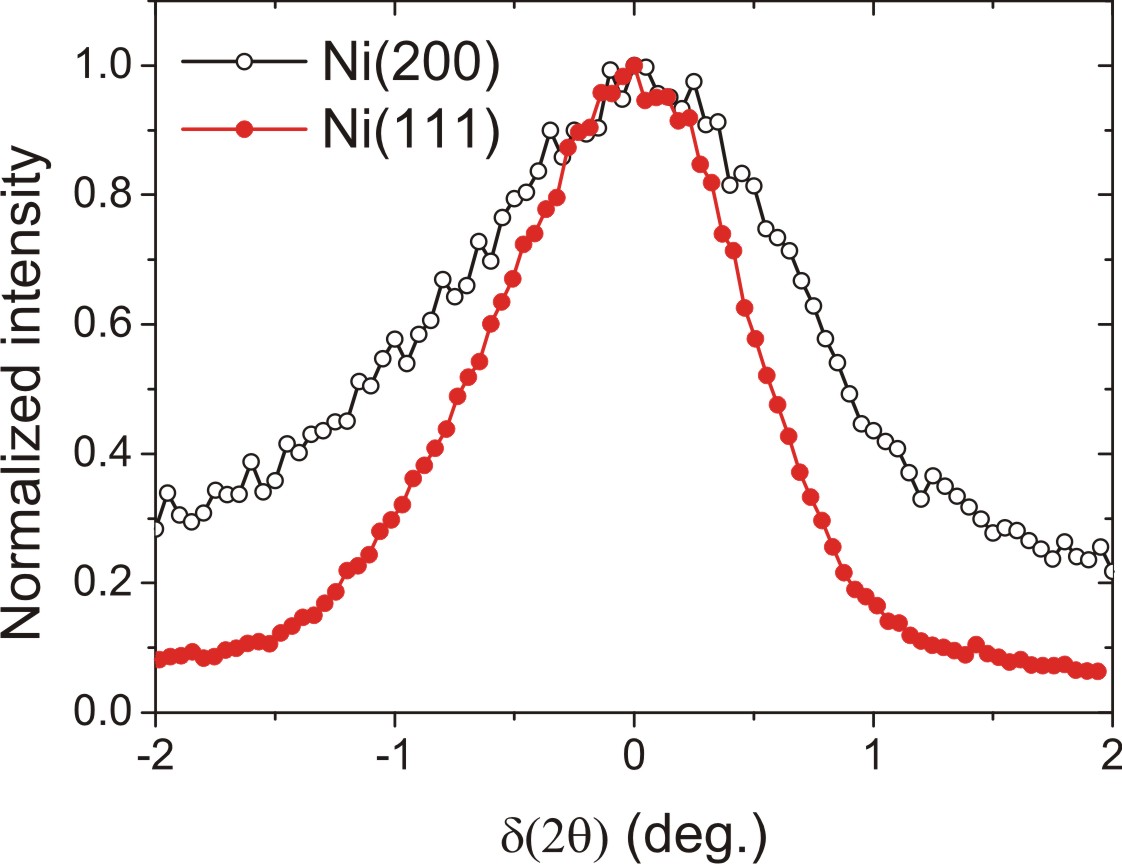}
\caption{XRD 2$\theta$-$\theta$ scan of Ni(200) and
Ni(111).}\label{fig:XRD_Ni(200)(111)}
\end{figure}

In principle, a uniformly magnetized single domain spherical
particle has no shape anisotropy, because the demagnetizing
factors are isotropic in all directions. However, in the case of a
nonspherical sample it will be easier to magnetize along a long
axis than along a short one. The FWHM of the Ni peak in XRD
2$\theta$-$\theta$ scans is a measure of the crystallite size.
Along the Ni[111] direction, the crystallite size is estimated to
be around 8 nm. For the in-plane direction, the crystallite size
could be estimated by measuring a diffraction plane not parallel
with sample surface. We chosen Ni(200) diffraction plane. Figure
\ref{fig:XRD_Ni(200)(111)} shows the comparison of the normalized
Ni(111) and (200) peaks. Actually, the Ni(200) is broader than
(111), which means the crystallite size along the in-plane
direction is even smaller than the [111] direction. Therefore, the
shape anisotropy is not the key reason to induce the easy axis
along the in-plane direction.

Now we consider the strain anisotropy. This kind of anisotropy is
often described by a magnetoelastic energy term
\begin{equation}\label{magnetoelastic}
    k_{magnetoelastic}=-\frac{3}{2}\lambda_{S}\sigma\cos^{2}{\phi}
\end{equation}
where $\lambda_{S}$ is the magnetostriction constant, $\sigma$ the
stress, and $\phi$ the angle between magnetization and the strain
tensor axis \cite{shikazumi}. The strain $\epsilon$ along [111]
direction of our Ni NCs can be calculated according to XRD
measurement.

The lattice spacing is given by Bragg law:
\begin{equation}\label{Bragg}
    2d_{exp}\sin\theta=\lambda
\end{equation}
where $d_{exp}$ is the lattice spacing, $\theta$ the Bragg angle,
and $\lambda$=0.154 nm the wavelength of x-ray. $\theta$ is
obtained from Figure \ref{fig:XRD_CoNiZnO_asimp}. The strain is
defined as following
\begin{equation}\label{strain}
    \epsilon=(d_{exp}-d)/d
\end{equation}
where $d$ is the theoretical lattice spacing for bulk Ni or Co.

\begin{table*} \caption{\label{tab:anisotropy}Comparison of magnetic-anisotropy energy density: magnetocrystalline $K_1$, magnetoelastic $K_{ME}$. $\lambda_{S}$ is the magnetostriction coefficient, $E$ the Young modulus, $\epsilon$ the elastic strain, and $\sigma$ the stress ($\sigma=E\epsilon$).}
\begin{ruledtabular}
\begin{tabular}{ccccccc}
  Nanocrystals & $k_{ME}$ (10$^3$ J$m^{-3}$) & $k_1$ (10$^3$ J$m^{-3}$)\footnotemark[1]& $\lambda_{S}$\footnotemark[1] & $E$ (GPa)\footnotemark[1] & $\epsilon$ & $\sigma$ (Gpa)\\
  \hline
  Ni (As-imp.) & 54  & -5.7 & -24$\times$10$^{-6}$ & 200 & 0.011 & 2.2 \\
  Ni (823 K ann.) & 50  & -5.7 & -24$\times$10$^{-6}$ & 200 & 0.010 & 2.0 \\
  Ni (923 K ann.) & 25  & -5.7 & -24$\times$10$^{-6}$ & 200 & 0.005 & 1.0 \\
  hcp-Co (As-imp.) & 125 & 500 & -5.5$\times$10$^{-5}$ & 209 & 0.011 & 2.3 \\
\end{tabular}
\end{ruledtabular}
\footnotetext[1]{Data from
Refs.\cite{kumar:064421,kazakova:184413}.}
\end{table*}

Using the approach and parameters in Refs.
\cite{kumar:064421,kazakova:184413}, the magnetoelastic anisotropy
constant is calculated (see table \ref{tab:anisotropy}). For Ni
NCs, the magnetoelastic anisotropy constant is one order higher
than the magnetocrystalline anisotropy constant. Therefore the
magnetoelastic anisotropy energy dominates the total anisotropy
energy. This finding demonstrates the possibility to tune the
magnetic properties by embedding magnetic NCs in different host
matrix. The annealing at 823 K does not change the strain status
significantly. After annealing up to 923 K, the elastic strain is
partially released. However, the magnetoelastic anisotropy energy
is still much higher than the magnetocrystalline one. Note the
anisotropy change in Figure \ref{fig:SQUID_NiZnO_ann}(d) after 923
K annealing by comparing with the 823 K annealing. There could be
another anisotropy source after 923 K annealing, which will be
discussed in section \ref{923K}.

For hcp-Co, the magnetocrystalline anisotropy constant is very
large, and it is difficult to be dominated by other anisotropy
energy contribution (as shown in Table \ref{tab:anisotropy}).
Therefore the hcp-Co NCs in the as-implanted state still persist
the bulk like anisotropy behavior. However, the annealing at 823 K
resulted in coexistence of fcc-Co and hcp-Co NCs. Since fcc-Co is
a meta-stable state, its magnetic and mechanic data are very
limited. Therefore in this paper, we will not discuss its
magnetoelastic anisotropy energy. Nevertheless fcc-Co[112] is the
easy axis rather than [111], which could be due the lattice
strain, as the same reason for Ni nanocrystals in ZnO. After
annealing at 923 K, hcp-Co is the only phase, and the elastic
strain is not significantly changed compared with the as-implanted
state (see Figure \ref{fig:XRD_CoNiZnO_ann}(a)). However, the
anisotropy is obviously changed. Like the Ni case, there could be
another anisotropy source, which will be discussed in section
\ref{923K}.

\subsection{Annealing at 923 K}\label{923K}
After annealing at 923 K, the magnetic properties, namely ZFC/FC
magnetization, and magnetic anisotropy, are changed significantly
compared with other samples. For the Co case, $T_B$ in the ZFC
curve increases drastically, while the FC curve is not
mono-decreased with increasing temperature. For Ni case, there is
a broad plateau in ZFC curve.

One explanation for the higher $T_B$ is the increase and
broadening of the Co crystallite size. Jacobson \etal~ calculated
the ZFC curves by varying different parameters, including the size
distribution \cite{jacobsohn:321}. It is found that a slight
broadening can result in a very broad and high ZFC curve. This can
well explain the ZFC magnetization, but not the FC magnetization.
Moreover the drastic changes in coercivity, and magnetic
anisotropy cannot be explained in such an approach. In addition,
Table \ref{tab:CoNiZnO_ann} lists the comparison of the XRD peak
area and the magnetization upon annealing. Note that the XRD peak
area decreases drastically by almost one order after 923 K
annealing, which indicates the decreasing of the amount of the Co
or Ni NCs, however the saturation magnetization only decreases
slightly.

\begin{figure} \center
\includegraphics[scale=0.9]{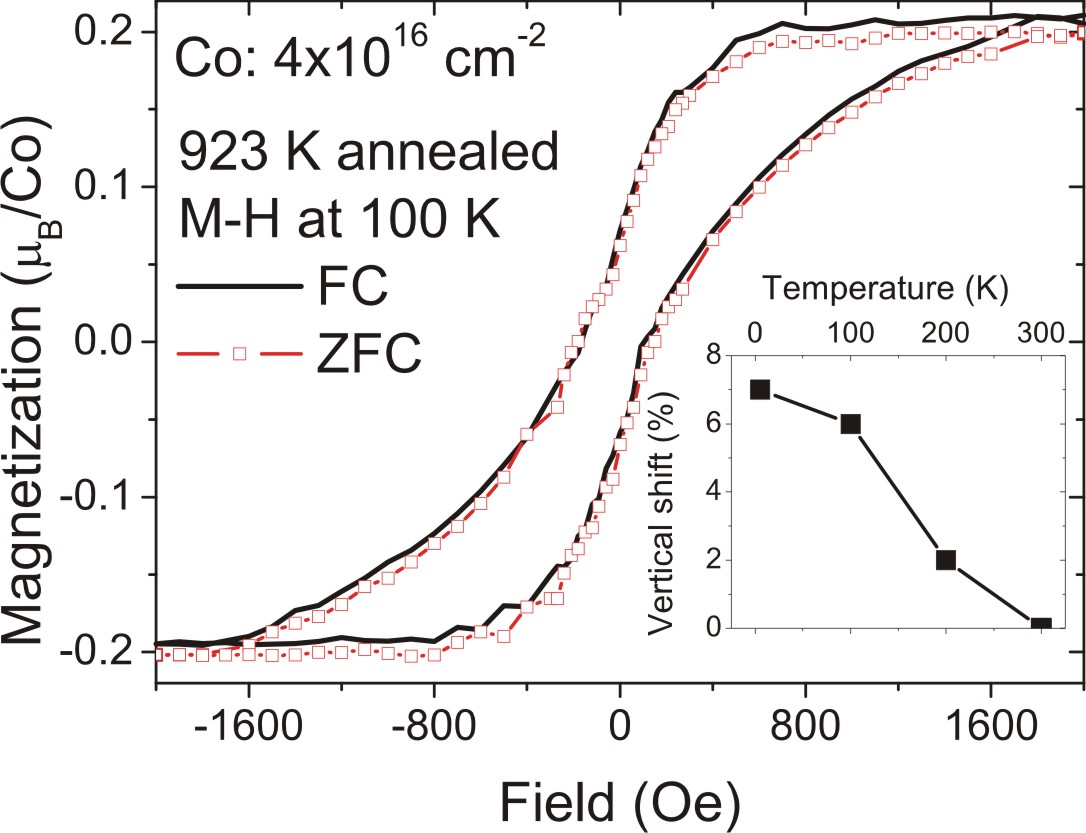}
\caption{The sample annealed at 923 K shows a vertical shift along
the magnetization axis after field cooling (H=2000 Oe) due to the
exchange coupling between FM and AFM materials. The field is along
ZnO[0001] direction. The inset shows the temperature dependence of
the vertical shift.}\label{fig:Co_MH_ZFCFC}
\end{figure}

Another explanation is due to other anisotropic energy
contribution. Skumryev \etal~have found that the exchange coupling
between Co NCs and their CoO shells drastically increase the
blocking temperature \cite{skumryev:850}. In our XRD measurement,
the amount of metallic Co and Ni decreases after annealing at 923
K, which is very probably due to the oxidation. Therefore we could
assume the formation of a Co/CoO core/shell structure in our
sample, and the exchange coupling increases the blocking
temperature. The exchange coupling is further confirmed by the
vertical shift of the magnetization loop (Figure
\ref{fig:Co_MH_ZFCFC}). The loop-curve is shifted along the
magnetization axis after cooling from 350 K in an applied field.

The vertical shift of the magnetization loop is a strong evidence
for the presence of an interfacial interaction between an
anitferromagnet (AFM) and a ferromagnet (FM) \cite{dobrynin05},
\ie~the hcp Co NCs are surrounded by CoO, which is amorphous, and
therefore cannot be detected by XRD. Dobrynin
\etal~\cite{dobrynin05} presented a model to discuss the exchange
coupling of nanoscaled Co/CoO core/shell structures. Below a
critical size (12 nm) of Co cores, the interfacial exchange energy
is larger than both the Zeeman energy of FM and the anisotropy
energy of the AFM due to a large surface-to-volume ratio of NCs,
and consequently some spins in the FM part can be frozen by the
AFM part, leading to a vertical shift along the magnetization axis
after field cooling. This model well explains our Co/CoO system
with an average diameter of 10 nm for Co NCs. The vertical shift
of the magnetization loop decreases with increasing temperature
and disappears at a temperature between 200 and 300 K. This is
consistent with the N\'{e}el temperature of 290 K for CoO
\cite{skumryev:850}. Such a vertical shift is also observed in the
FC-loop measured along
ZnO[10$\underline{1}$0]$\parallel$hcp-Co[10$\underline{1}$0]
(in-plane) (not shown).

For the Ni implanted sample annealed at 923 K, the hysteresis loop
was also measured under both ZFC and FC conditions (not shown). A
similar shift along the magnetization axis is observed. Therefore
we would assume the formation of a Co/CoO (Ni/NiO) core/shell
structure. The exchange coupling between Co (Ni) and CoO (NiO)
contributes another anisotropy energy. This explains the high
blocking temperature and the change in magnetic-anisotropy after
annealing at 923 K.

Note that for both of cases, the magnetization is only slightly
decreased after annealing at 923 K. This is controversial with the
drastic decrease of the diffraction intensity (see Figure
\ref{fig:XRD_CoNiZnO_ann}). One possible reason is the formation
of small Co or Ni nanocrystals due to annealing. As shown in Table
\ref{tab:CoNiZnO_asimp}, only around 10-40\% of Co or Ni form as
metallic nanocrystals and others remain as dispersed ones.
Annealing at 923 K, on one hand, oxidized some metallic
nanocrystals partially, on the other hand, could induce the
gettering of dispersed Co and Ni and result in small metallic
nanocrystals. Below a critical size, nanocrystals are
non-detectable by XRD. Another reason could be that only a thin
shell of Co or Ni nanocrystals transforms into oxides, which
results in the invisibility of oxides in XRD even they are
crystalline. Obviously a detailed investigation using transmission
electron microscopy or other more sensitive techniques should be
performed to clarify this controversy.

\section{Conclusions}\label{conclusion}
A thorough characterization of the structural and magnetic
properties has been presented on Co and Ni implanted ZnO single
crystals. The results by SR-XRD and SQUID magnetometery correlate
well with each other. The major conclusions are summarized as
follows.

(1) Co or Ni NCs have been formed in ZnO upon ion implantation.
Their crystalline sizes, generally below 10 nm, increase with
increasing fluence. From several to 37\% percent of implanted Co
or Ni is in metallic states, while the remaining could be diluted
into ZnO matrix. The Co or Ni NCs are the origin of the measured
ferromagnetism.

(2) The Co and Ni NCs are crystallographically oriented with
respect to ZnO host matrix. The orientation relationship is as
follows:
hcp-Co(0001)[10$\underline{1}$0]$\parallel$ZnO(0001)[10$\underline{1}$0],
and Ni(111)[112]$\parallel$ZnO(0001)[10$\underline{1}$0]. This
well ordered structure of NCs could result in a rather smooth
interface between them and ZnO host, and makes the hybrid of
ferromagnetic NCs and semiconductors promising for spintronics
functionality.

(3) Magnetic anisotropy is observed for Co or Ni NCs in ZnO.
Especially for the Ni NCs, the anisotropy is different from the
bulk crystals. The extra anisotropy energy is attributed to the
lattice strain impressed from the host matrix. This opens a route
to artificially tune the magnetic properties of nanoparticles by
selecting of substrate materials.

(4) The structure and magnetic properties of Co or Ni NCs embedded
inside ZnO can be tuned by post-annealing. For the Co case, 823 K
annealing results in the co-exists of fcc-Co and hcp-Co. The
magnetic anisotropy is changed from out-of-plane to in-plane.
Annealing at 923 K could have partially oxidized metallic Co and
Ni, and result in Co/CoO (Ni/NiO) core/shell structures. After
annealing at 1073 K, no Co or Ni NCs can be detectable within the
detection limit of SR-XRD, at the same time, the samples shows no
pronounced ferromagnetism down to 5 K.

\pagebreak

\end{document}